# Tailoring Thermal Conductivity of Single-stranded Carbon-chain Polymers through Atomic Mass Modification

Quanwen Liao[1], Lingping Zeng[2], Zhichun Liu[1,*], Wei Liu[1,*]

1. School of Energy and Power Engineering, Huazhong University of Science and Technology (HUST), Wuhan 430074, China

2. Department of Mechanical Engineering, Massachusetts Institute of Technology, Cambridge, Massachusetts 02139, USA

* Corresponding authors. E-mail: zcliu@hust.edu.cn (Zhichun Liu), w_liu@hust.edu.cn (Wei Liu)

# ABSTRACT

Tailoring the thermal conductivity of polymers is central to enlarge their applications in the thermal management of flexible integrated circuits. Progress has been made over the past decade by fabricating materials with various nanostructures, but a clear relationship between various functional groups and thermal properties of polymers remains to be established. Here, we numerically study the thermal conductivity of single-stranded carbon-chain polymers with multiple substituents of hydrogen atoms through atomic mass modification. We find that their thermal conductivity can be tuned by atomic mass modifications as revealed through molecular dynamics simulations. The simulation results suggest that heavy homogeneous substituents do not assist heat transport and trace amounts of heavy substituents can in fact hinder heat transport substantially. Our analysis indicates that carbon chain has the biggest contribution (over 80%) to the thermal conduction in single-stranded carbon-chain polymers. We further demonstrate that atomic mass modifications influence the phonon bands of bonding carbon atoms, and the discrepancies of phonon bands between carbon atoms are responsible for the remarkable drops in thermal conductivity and large thermal resistances in carbon chains. Our study provides fundamental insight into how to tailor the thermal conductivity of polymers through variable substituents.

## Introduction

Polymers have been widely utilized in a wide variety of technological applications due to their outstanding physical properties, such as high toughness, low density and cost, and high corrosion resistance. One drawback of polymers, though, is their poor heat conduction abilities that significantly limit their applications.[1-4] The ultra-low thermal conductivities ($\kappa$) of bulk polymers, on the order of 0.1 Wm$^{-1}$K$^{-1}$ at room temperature[5], mainly results from randomly twisting chain structures that cause a multitude of phonon scatterings inside the polymers and consequently shorten the heat-carrying phonon mean free paths[6,7]. Enhancing polymers' thermal conductivity well beyond the amorphous level has been of extensive research interest recently.[8-11]

Generally, there are two approaches to enhancing polymers' thermal conductivities: dopant-based and polymer-based thermal conduction enhancement. Doping the originally thermally insulating polymer matrix with a small percentage of thermally conductive additives, such as graphene[12-14], carbon nanotubes (CNTs)[15-20] or metal powders[21], to form polymer composites is a common practice to improve the polymers' heat conduction ability. For example, Marconnet *et al.*[19] systematically studied the effect of carbon nanotube density on improving the thermal conductivity of an epoxy and observed that introducing ~ 17% volume fraction of aligned multi-walled carbon nanotubes can enhance the epoxy's thermal conductivity by a factor of 18. Some of co-authors' previous work also showed that the ideal doping model can significantly improve the heat transfer performance of CNT-PE composites.[22] However, experimentally the poor filler dispersion and high interfacial thermal resistance between the filler and the polymer matrix significantly limit the practically achievable level of enhancements.[12,23,24] Moreover, the additives may be harmful to polymers' outstanding physical properties.

The second approach, termed here as polymer-based thermal conduction enhancement, was pioneered by Chen group's work on crystalizing polymer structure domain to improve its thermal conductivity.[8,9] The basic idea is to reduce the phonon scattering sites by reducing the number of defects by fabricating polymer samples with well-aligned structures. Their numerical and experimental observations of remarkably enhanced polymer thermal conductivities provided fundamental guidance on how to design and fabricate polymers with high heat transfer ability. Several subsequent works showed that increasing the molecular chains' orientation order in amorphous polymers can greatly increase the thermal conductivities of polymers through mechanical strains and nanoscale templates.[3,9,11,25,26] For example, one recent experimental study revealed that pure polythiophene nanofibres can have a thermal conductivity up to ~4.4 $Wm^{-1}K^{-1}$ (more than 20 times higher than the bulk polymer value) while remaining amorphous.[11] Kim et al. showed that a blend of two polymers with high miscibility and appropriately chosen linker structure can yield a dense and homogeneously distributed thermal network, and therefore a significant increase in cross-plane thermal conductivity was observed, leading to a thermal conductivity larger than 1.5 $Wm^{-1}K^{-1}$.[27] Luo et al. studied the thermal conductivities of various polymer nanofibers that are found to depend greatly upon different backbones.[4] These work paved the way to improve the thermal conductivity of polymers for a range of potential applications.

Despite the significant progress made on improving polymer's thermal conductivity, the impact of different functional groups on polymer thermal conductivity is still an open question. In this work, we numerically study the thermal conductivities of single-stranded carbon-chain polymers (SCP) using equilibrium molecular dynamics (EMD) and the thermal resistances of modified single polyethylene chain (MSPEC) using classical non-equilibrium molecular dynamics (NEMD) in detail. Both SCP and MSPEC are derived from single polyethylene chain (SPEC) via atomic substitutions, namely, SPEC's hydrogen atoms are replaced by substituent atoms or functional groups, such as halogens, hydroxyls, carboxyls, phenolic groups, *etc*. The MSPEC is a SPEC with a functional group modification at its middle site. To make the modeling of variable substituents tractable, the hydrogen atoms are modified to different masses instead of being replaced by actual substituents, and the bonding strength between carbon and modified hydrogen atoms remains the same. In addition, we neglect the impact of possible changes of molecular structures, such as possible segment long-chain disorders caused by cis-trans isomerism, due to actual substituents as well. Figure 1 shows the ratio of the depth of potential well $k_{IJ}$ to the relative atomic mass *m* versus *m* under the Universal force field[28,29] (UFF). The good agreement between mass modified cases (open squares) and real elements (solid dots) indicates that these modified masses correspond to certain elements in nature, or at least within reasonable mass range. With these simplifications, our simulation results provide general insight into the fundamental mechanism of tailoring polymer thermal conductivity via atomic mass modification. In the following section, we first describe the modeling framework and simulation procedures based on molecular dynamics and then proceed to discuss the simulation results and analyze the mechanism behind thermal conductivity control. The simulation results explicitly show that the thermal conductivity of SCP can be tuned by variable substituents of hydrogen atoms. Our study helps gain fundamental insight into how to improve the flexibility of controlling polymer thermal conductivity and may inspire an array of polymer-related applications.

**Results and Discussions**

As shown in Fig. 2, two representative models of SCP, Model C-X and Model C-$H_mY_n$, are applied to study the functionalized polymer thermal conductivity. In Fig. 2, X atoms and Y atoms denote the modified hydrogen (H) atoms, and *m* and *n* denote the ratio between the number of hydrogen atoms and the number of Y atoms. To begin with, we investigate the size effect in the thermal conductivity of SPEC as a function of chain length at room temperature. As shown in Fig. 2a, the thermal conductivity displays a very weak dependence on the chain length above 10 nm, indicating that a chain length of 10 nm is sufficiently long for us to neglect the size effect. Consequently, we use a chain length of 10 nm for the Model C-X and Model C-$H_mY_n$ in the subsequent simulations.

In Model C-X, all the H atoms are substituted by X atoms in SPEC. The values used for atomic masses of X atoms ($m_X$) range from 0.5 to 127 g/mol to being consistent with the range of real atomic masses. Figure 2b shows that the thermal conductivity of Model C-X has a strong dependence on the atomic masses of X atoms and decreases with increasing $m_X$, indicating that light atoms can facilitate heat conduction in SCP. The best power-law fit gives $\kappa \sim m_X^\delta$ with $\delta$ = -0.21 ± 0.02. In order to enhance the thermal conduction in carbon-chain polymers, the heavy substituents of H atoms should be consciously avoided in Model C-X.

In C-$H_mY_n$, a portion of H atoms are substituted by Y atoms uniformly. In other words, H atoms are replaced by Y atoms at equal spacing along SPEC. We define the fraction of H atoms ($\Phi_H$) as the percentage of the number of H atoms with respect to the total number of H atoms and Y atoms. The values used for $m_Y$ are 19, 35.5 and 80 g/mol. When $\Phi_H$ = 100% (or $n$ = 0), Model C-$H_mY_n$ reduces to SPEC; and when $\Phi_H$ = 0 (or $m$ = 0), Model C-$H_mY_n$ reduces to Model C-X. Figure 2c shows thermal conductivity of Model C-$H_mY_n$ as a function of $\Phi_H$. As we can see, SPEC has the largest thermal conductivity, and generally the Model C-$H_mY_n$ (m ≠ 0, n ≠ 0) with heterogeneous substituents (H atoms and Y atoms coexist) possesses lower thermal conductivity than SPEC and Model C-X that have homogeneous substituents (only H atoms or X atoms exists). Note that substituting hydrogen atoms by a very small fraction ($\Phi_H$ = 96.88%) will dramatically reduce the thermal conductivity of SPEC, by approximately 74.4%. These results imply that heavy homogeneous substituents in Model C-X do not assist thermal transport and even trace amounts of heavy substituents in Model C-$H_mY_n$ can hinder heat transport substantially. Below we analyze the phonon modes to understand these thermal conduction behaviors in SCP.

Figure 3 shows the results of phonon mode analysis for Model C-X. In the simulations, we calculate the velocity auto-correlation functions (VACF) of C atoms and X atoms separately. Then their phonon power spectrums are extracted from VACF through Fast Fourier Transform. Figure 3b shows the normalized power spectrum of X atoms. It is clear that the phonon modes of X atoms greatly depend on $m_X$. As $m_X$ increases, the dominant heat-carrying phonon modes gradually shift from high frequencies (around 120 THz) to relatively low-frequency region (0 ~ 10 THz). This redshift in full spectra and reduction in bandwidth are also present explicitly in Fig. 3d that shows the corresponding cumulative power spectrum of X atom. Clearly, the cumulative distribution curves shift to the left as $m_X$ increases, indicating higher contribution from low-frequency phonon modes. Therefore, we attribute the change in thermal conductivity of Model C-X to originate from the changes in the heat-carrying phonon modes in X atoms. Due to the covalent bonds between C atoms and X atoms, as the phonon bands of X atoms change with $m_X$, the phonon bands of C atoms also experience a significant variation. Figures 3a and 3c show the normalized power spectrum and corresponding cumulative power spectrum of C atoms. As illustrated by Fig. 3a, the contribution of low-frequency region (0 ~ 15 THz) to thermal transport is gradually weakened as $m_X$ increases, and only a few phonon modes survive when $m_X$ reaches 127 g/mol. Another feature worth noting is that the contribution of mid-frequency phonon modes (15 ~ 45 THz) rises while that of the low-frequency region is reduced. This phenomenon can be clearly seen in Fig. 3c as well. As the $m_X$ increases, cumulative power curves become flatter at low-frequency region and steeper at mid-frequency region.

There is a direct relationship between variation of carbon chain's phonon bands and thermal conduction behaviors of SCP. As we know, phonons represent the quantized normal modes of lattice vibrations that transfer heat in a material. The phonon transport behaviors in the system define the thermal conduction properties of materials. Generally, low-frequency phonons play a predominant role in thermal conduction due to their high group velocities and long mean free paths. The modifications of X atoms lower the phonon frequency of X atoms in full spectra. The resulting low-frequency phonons of X atoms scatter low-frequency phonons in carbon chain. Therefore, the thermal conductivity of Model C-X decreases with increasing $m_X$. As for Model C-$H_mY_n$, heterogeneous substituents lead to conflicting phonon bands of C atoms in carbon chain, so thermal conduction could be blocked dramatically by only a tiny percentage of substituents. These variations of carbon chain's phonon bands partially explain the thermal conduction behaviors of Model C-$H_mY_n$. However, to make these mechanisms more persuasive and more profound, we should analyze the contributions of carbon chains to thermal conduction in SCP and the detailed thermal resistances caused by phonon band discrepancies of heterogeneous substituents as well.

Finally, we use NEMD simulations to study thermal conduction of carbon chains in MSPEC as shown in Fig. 4. The middle hydrogen atom (red atom) is modified to variable masses. The fixed boundary condition is applied in the longitudinal direction. The simulation system is divided into 40 slabs.[30] The temperatures of the heat source and heat sink are set at 310 K and 290 K, respectively. The open circles show the temperature profile of carbon chains for $m^* = 25$ g/mol. We fit the linear parts to extract the temperature jump ($\Delta T$) at atomic modification sites. The thermal resistance $R$ is calculated through the equation $R = \Delta T / J$, where $J$ is heat flux.

Figures 5a and 5b show the results of thermal resistances as a function of modified masses in MSPEC. The heat flux ratio of carbon chains ($J_{CC}$ *ratio*) is shown as solid squares in Fig. 5a, where the contributions of carbon chains are over 80% in the total heat flux. This high heat flux contribution indicates that carbon chains contribute predominantly to the heat conduction in SCP. We only focus on the thermal resistance of carbon chains in order to eliminate the temperature fluctuations of hydrogen atoms. The solid circles show thermal resistances of carbon chains at modification sites ($R_{CC}$) obtained by using $J_{CC}$ and $\Delta T$. The $R_{CC}$ increases first with increasing modified mass and then it saturates at an approximately constant value (0.7 K/nW). $R_{CC}$ approaches its convergence value at about $m^* = 12$ g/mol. These results demonstrate that mass modifications of hydrogen atom can hinder heat conduction of carbon chains significantly, consistent with the results of Model C-$H_mY_n$. From Fig. 1, we can see that all these presented real elements have a value of $m$ larger than 12 g/mol. This suggests that heterogeneous substituents will hinder heat conduction of SCP greatly in practical systems as well. Figure 5b illustrates the heat fluxes ($J_{CC}$) and $\Delta T$ of carbon chains versus $m^*$. When $m^* < 20$ g/mol, $J_{CC}$ decreases and $\Delta T$ increases rapidly with increasing $m^*$. When $m^* > 20$ g/mol, both $J_{CC}$ and $\Delta T$ show converging behavior, to about 3.5nW and 2.6K.

Figures 5c and 5d show the corresponding phonon mode analyses of MSPEC. Figure 5c shows the cumulative normalized power spectra of the carbon atoms, which are bonded to the modified hydrogen atom. With increasing $m^*$, the contribution of low-frequency phonons (0 ~ 20 THz) decreases as denoted by the orange arrow, though these curves have some irregular behaviors at 0 ~ 1 THz. We believe these irregular behaviors are attributed to the long mean free paths of low-frequency phonons. Above all, the phonon band differences between MSPEC ($m^* \neq 1$ g/mol) and SPEC ($m^* = 1$ g/mol) explains the large $R_{CC}$ at modified sites. To quantify the phonon band differences, we use the normalized overlap of phonon bands of SPEC and MSPEC calculated by

$$S_i = \frac{\int_0^\infty \min[P_{m^*=1}(\omega), P_{m^*=i}(\omega)] d\omega}{\int_0^\infty P_{m^*=1}(\omega) d\omega}, \qquad (1)$$

where $\omega$ is the frequency, $P_{m^*=i}(\omega)$ is the power spectrum of carbon atom at modified site when $m^* = i$ g/mol, $S$ represents the overall overlap of phonon bands, denoting phonon bands' coherence of carbon atoms at modification sites. Figure 5d shows the results of overall overlap of phonon bands versus $m^*$. We can see that mass modifications make $S$ plunge considerably (by more than 0.5), and $S$ continues to slide as $m^*$ increases. This indicates that mass modifications decrease the phonon bands' coherence at modified site significantly, and thus a large thermal resistance is introduced. The phonon mode analyses explain the trend of $R_{CC}$ in MSPEC very well. Based on our analyses, a significant increase in Fig. 2c when $\Phi_H$ = 50% results from good phonon bands' coherence, because each carbon atoms is bonded to both a Y atom and an H atom.

To justify our results, we calculated the thermal conductivity of single PE, polyvinyl alcohol (PVA) and polypropylene (PP) chains employing COMPASS potential[31] through EMD. There are no atomic mass modification and constant depth of potential well here, so it describes more accurately the real polymer chains than Model C-X. Figure 6 shows the thermal conductivity of single PE, PVA and PP chains with respect to length. We can see that their thermal conductivities are not sensitive to length. The thermal conductivity of PE described by COMPASS potential is consistent with the results using AIREBO potential. And the single PVA and PP chains with a length of 160nm have a thermal conductivity of 7.01 and 5.83 $Wm^{-1}K^{-1}$, respectively. PVA and PP possess very similar thermal conductivity, but their thermal conductivity is far smaller than PE's. It agrees well with the results of Model C-X and Model C-$H_mY_n$. The similar thermal conductivity of single PVA and PP chains results from similar atomic mass of hydroxyl and methyl. However, the substituents at fifty percentage of carbon atoms introduce phonon band differences within carbon backbones, which make the thermal conductivity of PVA and PP much smaller than the thermal conductivity of Model C-X with atomic mass modifications of 17 and 12 g/mol.

**Conclusions**

We numerically study the thermal conductivities of single-stranded carbon-chain polymers (SCP) using AIREBO potential by EMD and the thermal resistances of modified SPEC (MSPEC) by NEMD. In our simulations, a simplification of substituents of hydrogen atom is applied, and such simplifications are shown to be effective. EMD simulations reveal that homogeneous substituents in Model C-X can diminish its thermal conductivity significantly, and trace amounts of substituents ($\Phi_H$ = 96.88%) in Model C-$H_mY_n$ can substantially hinder heat conduction (by about 74.4%). In addition, NEMD simulations show that carbon chain has the biggest contribution (over 80%) to the total heat flux in SCP, and heterogeneous substituents introduce a large thermal resistance at substituting sites. The results of NEMD simulations are consistent with those of EMD simulations. Analyses of phonon power spectra demonstrate that mass modifications of hydrogen atoms influence the phonon bands of carbon atoms that bonded to them. For homogeneous substituents (Model C-X), the redshift of phonon bands accounts for the reduction of thermal conductivity; for heterogeneous substituents (Model C-$H_mY_n$) and MSPEC, the differences of phonon bands between carbon atoms are responsible for the striking drops in thermal conductivity and large thermal resistances in carbon chains. The simulations of single PE, PVA and PP chains employing COMPASS potential justify our results. Most importantly, the intrinsic thermal conductivity of SCP can be tuned by variable substituents of hydrogen atoms. Our study may inspire manufacturing SCP of a specific thermal conductivity for a wide variety of applications.

**Methods**

Equilibrium molecular dynamics[32] simulations are used to study the thermal conductivity of SCP with a length of 10 nm at room temperature. The periodic boundary condition is applied in the longitudinal direction and the fixed boundary conditions are applied in the transversal directions. Figure 2 shows two representative models of SCP, Model C-X and Model C-$H_mY_n$, where $m$ and $n$ are integer variables. Thermal conductivity is derived from the Green-Kubo formula.[33] The values of thermal conductivity are extracted from six realizations with various initial velocity conditions. The EMD method has been detailed in the supplementary of Ref. [32].

Classical non-equilibrium molecular dynamics simulations are used to study the carbon chain's thermal resistances of MSPEC with a length of 10 nm as shown in Fig. 4. The middle hydrogen atom (denoted in red) is modified to variable masses ($m^* = 1 \sim 280$ g/mol). Figure 4 shows a typical setup and the corresponding temperature profile. The NEMD method has been detailed in the Ref. [34].

All simulations are performed by the large-scale atomic/molecular massively parallel simulator (LAMMPS) package.[35] Due to the simplifications mentioned in the introduction, both MSPEC and SCP are SPEC with atomic mass modifications. Thus the atomic interactions of both MSPEC and SCP are described by an adaptive intermolecular reactive empirical bond order (AIREBO) potential,[36] which is developed from the second-generation Brenner potential.[37] Additionally, an 8.5 Å cutoff distance is used for the long-ranged 12-6 Lennard-Jones interaction. The equations of motion are integrated by the velocity Verlet algorithm with a time step of 0.2 fs.

## ASSOCIATED CONTENT

### Acknowledgments

This project was supported in part by the National Natural Science Foundation of China: 51376069 (ZL) and the Major State Basic Research Development Program of China No. 2013CB228302 (ZL). The work was performed at the National Supercomputer Center in Tianjin and the calculations were performed on TianHe-1(A).

### Author contributions

### Additional Information

Competing financial interests: The authors declare no competing financial interests.

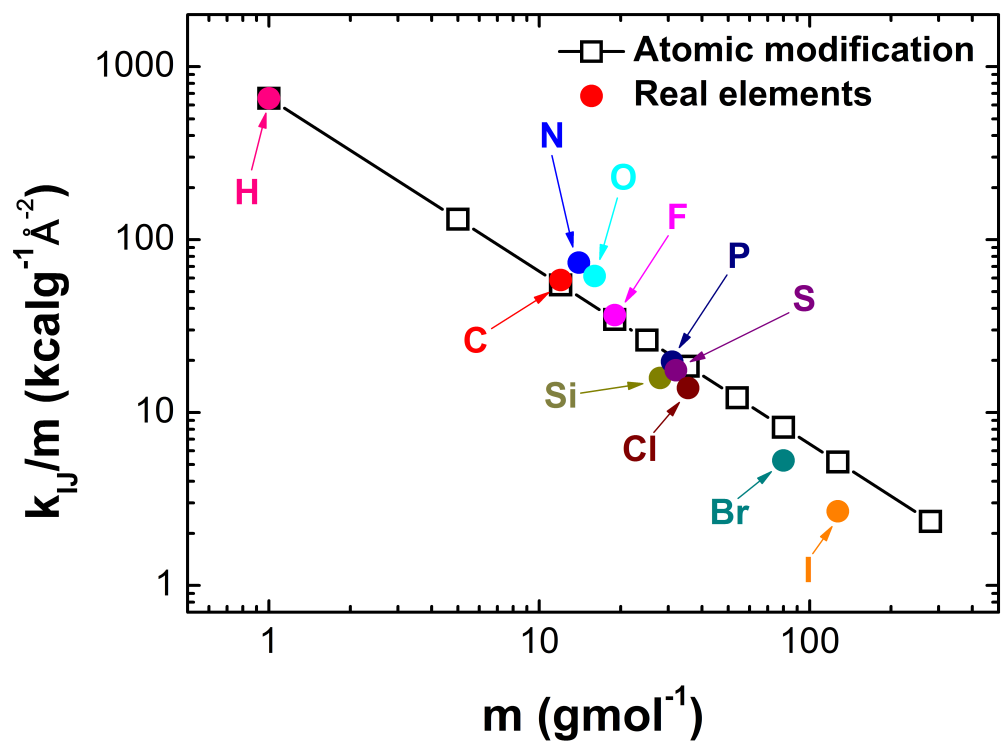

Figure 1. The depth of potential well per unit mass ($k_{IJ}/m$) versus relative atomic mass ($m$) under the Universal force field (UFF): $k_{IJ}$ is the force constant in units of (kcal/mol)/ Å$^2$, $m$ is the relative atomic mass in units of (g/mol). The solid dots represent covalent bonds between tetrahedral carbon and main group elements. The open squares represent covalent bonds between tetrahedral carbon and hydrogen atoms with mass modifications, where $k_{IJ}$ is fixed as the value of single C-H bond.

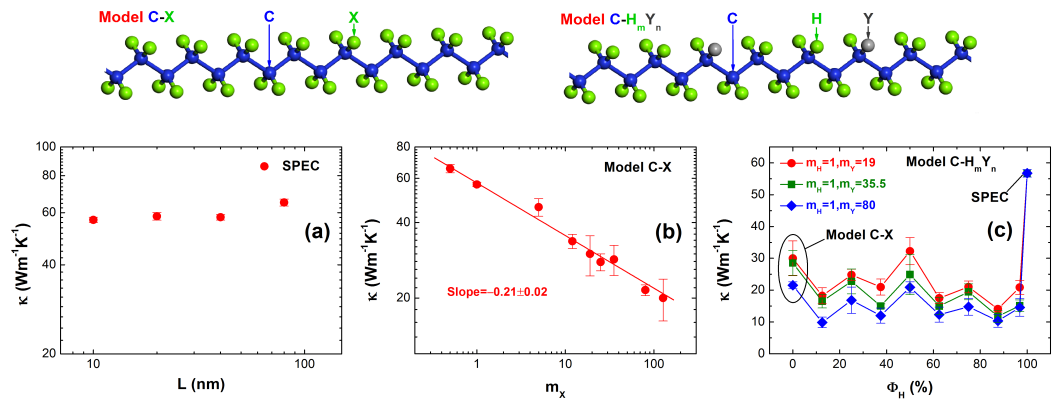

Figure 2. The representative models and thermal conductivity ($\kappa$) of SCP with different atomic masses of X atoms ($m_X$) calculated by equilibrium molecular dynamics (EMD) simulations at room temperature. (a) The thermal conductivity of SPEC ($m_X$ = 1 g/mol) versus chain length ($L$). (b) The thermal conductivity of Model C-X with varied $m_X$. (c) The thermal conductivity of Model C-$H_m Y_n$ with variable fraction of H atoms $\Phi_H$, where $\Phi_H$ is defined as the percentage of H atoms in the total H atoms and Y atoms.

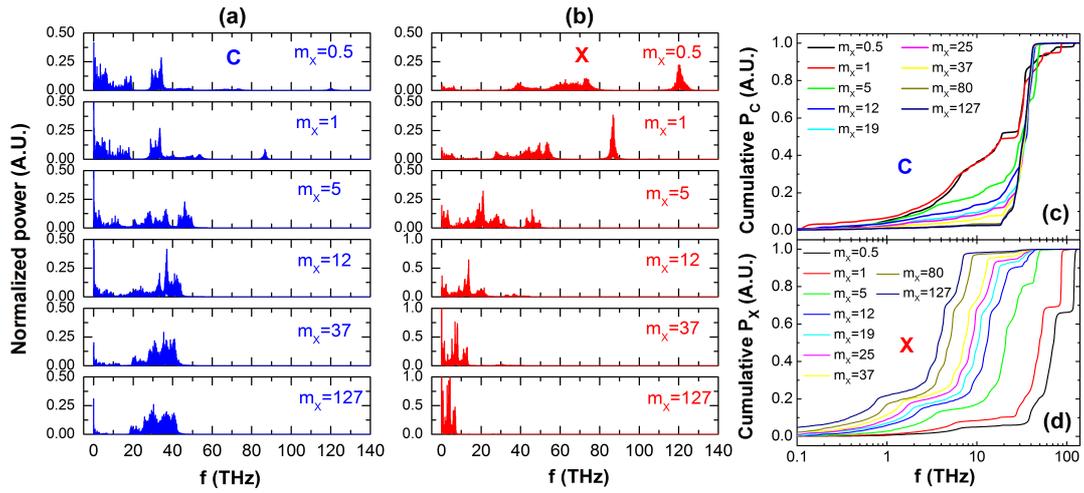

Figure 3. The phonon mode analyses of Model C-X with different $m_X$. (a) and (b) are the normalized power spectrum of C atoms and X atoms, respectively. (c) and (d) are the cumulative normalized power spectrum of C atoms and X atoms, respectively.

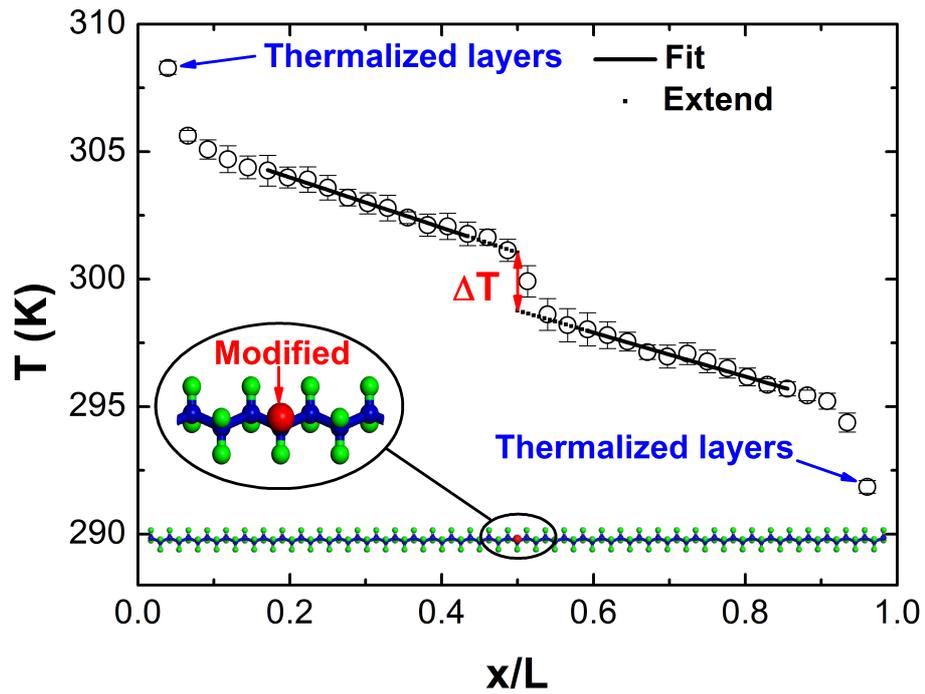

Figure 4. The typical setup and the corresponding carbon-chain temperature profile from NEMD simulations to extract thermal resistances. The middle hydrogen atom (red atom) is modified to variable masses in a MSPEC with a length of 10 nm.

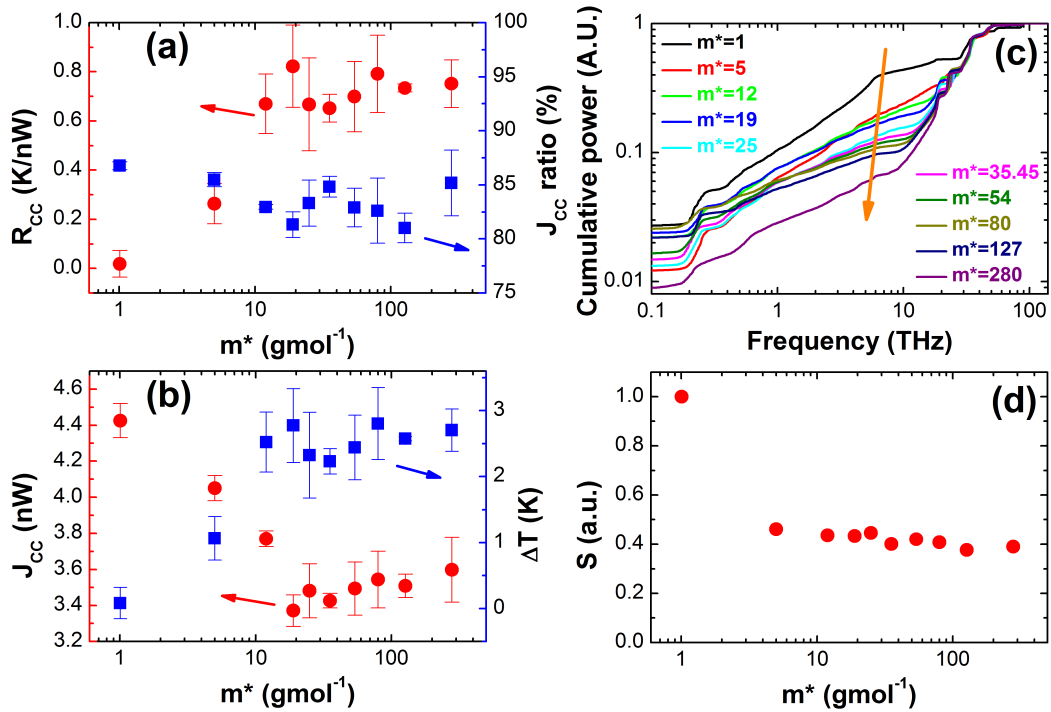

Figure 5. The thermal resistance analyses of MSPEC's carbon chains using NEMD simulations. (a) Thermal resistance (left) of carbon chain at modified hydrogen atoms and heat flux ratio (right) of carbon chains in total heat flux versus modified masses ($m^*$). (b) Carbon chain's heat fluxes and carbon chain's temperature jump ($\Delta T$) at modified hydrogen atoms versus $m^*$. (c) Cumulative normalized power spectrum of the carbon atom, which is bonded to modified hydrogen atoms. (d) The normalized overlap of the phonon bands ($S$) of SPEC and MSPEC versus $m^*$.

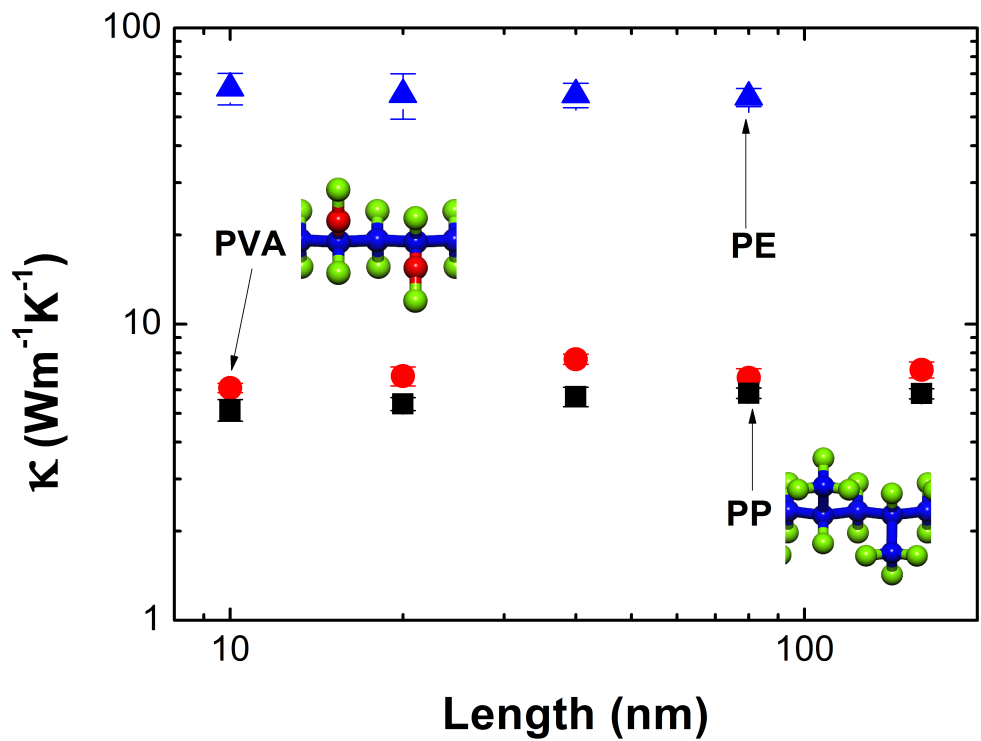

Figure 6. The thermal conductivity of single PE, polyvinyl alcohol (PVA) and polypropylene (PP) chains with respect to length using COMPASS potential through EMD.